\documentstyle[aps,preprint,12pt]{revtex}  
\begin{document}                
\title{Binding energy of negatively charged exciton in a semiconductor
quantum well: the role of interface defects}
\author{Luis C. O. Dacal$^{1,2}$, R. Ferreira$^1$, G. Bastard$^1$ and Jos\'{e} A. Brum$^{2,3}$}
\address{(1) Laboratoire de Physique de la Mati\`{e}re Condens\'{e}e  ENS , 24, rue Lhomond, 75005
-Paris, France}
\address{(2) IFGW-DFMC, Universidade Estadual de Campinas, C.P. 6165,
13083-970, Campinas-SP, Brazil}
\address{(3) Laborat\'{o}rio Nacional de Luz S\'{i}ncrotron - ABTLuS, C. P. 6192, 13084-971,
 Campinas-SP, Brazil}
%
\maketitle

\begin{abstract}                
We present a model to take into account the interface defects
contribution on the binding energy of charged exciton in
GaAs/Al$_{0.3}$Ga$_{0.7}$As quantum wells. The dependence of the
binding energy gain and of the trion size on the quantum well
width are variationally calculated. We show that the trion is more
sensitive to interface defects than the exciton.
\end{abstract}

\begin{center}


Corresponding author : Luis Carlos Ogando Dacal

Address : GPO-DFMC-IFGW, Universidade Estadual de Campinas, C.P. 6165\\
13083-970, Campinas (SP), Brazil

Phone: (005519)3788-5505

Fax: (005519)3289-3137

e-mail: ogando01$@$ifi.unicamp.br

\end{center}

\newpage
\section{Introduction}               

When low power is used to excite an intrinsic quantum well (QW),
its photoluminescence spectrum is dominated by an exciton which is
analogous to the H atom in semiconductor physics. This complex is
formed by the Coulomb attraction between one electron and one
hole. In the case of lightly modulation doped samples, the
presence of extra carriers inside the QW gives rise to new bound
states through the attraction of an extra electron (n-doped) or
hole (p-doped) by the excitonic electrical dipole forming a
charged complex, the trion, which may be negatively charged,
X$^-$, or positively charged, X$^+$, respectively. Lampert
\cite{Lampert58} was the first author to propose the stability of
such charged complexes in semiconductors. However, the first
experimental observation was only possible in QWs and it was made
by Kheng \textit{et al}.\cite{Kheng93}. The influence of carrier
localization potentials on the trion experimental observations is
an interesting and open question. Some experimental works show
evidences of carrier localization \cite{Eytan98} while other
results indicate that the trion is composed by free carriers
\cite{Ciulin96}. Moreover, the measured binding energies are
consistently higher than the theoretical ones, indicating the
possibility of trapped electrons. To shed some light on this
question, we variationally calculated the binding energies of
charged exciton in GaAs/Al$_{0.3}$Ga$_{0.7}$As QWs including the
presence of structural defects represented by attractive gaussian
potentials. This kind of defect is always present at the QW
interface due to the interdiffusion of well and barrier materials
during the QW growth process.

\section{Model}

We describe the QW, a GaAs layer between two
Al$_{0.3}$Ga$_{0.7}$As layers, using the effective mass and
envelope function frameworks. We start with the assumption that
the QW confinement is strong enough and that the interface defects
are weak enough to make reasonable the use of the noninteracting
electron and hole ground states in ideal QWs as the z-part (growth
direction) of the one particle trial wavefunction. The axial
symmetry will be preserved by the defect potential. This leads us
to describe the X$^-$ in-plane motion in terms of a center of mass
($\vec{R}$) and relative coordinates considering the following
intuitive picture : the X$^-$ is composed by an exciton and a
distant electron bound to its electrical dipole\cite{Peeters2000}.
Consequently, the relative coordinates are given by one electron
relative to the hole ($\vec{\rho}_1$) and the other one relative
the center of mass (CM) of this particle ($\vec{\rho}_2$) in the
X$^-$ case. We assume parabolic energy dispersions, so the
equivalent X$^+$ coordinates are obtained through the interchange
between electrons and holes.

The two electrons (holes in the case of X$^{+}$)
indistinguishability leads us to use a Slater determinant as
basis. We assume that the internal degrees of freedom are not
strongly affected by the defect
potential\cite{Bastard84,Heller97}, in other words, its main
effect is the localization of the CM, which is weakly coupled to
the internal dynamics. In the absence of structural defects and
external fields, only the singlet trion state with total angular
momentum in the z direction equal to zero is a bound state.
Therefore, we consider only this configuration for the orbital
part of the charged exciton trial wavefunction :

\begin{eqnarray}
\Psi_{0}=N_{i,j,m}.\chi_{0}(z_{h}).\chi_{0}(z_{e1}).\chi_{0}(z_{e2}).\phi_{m}^{0}(\vec{R})\times\left[\phi_{i}^{0}(\vec{\rho}_{1})
\phi_{j}^{0}(\vec{\rho}_{2})+\phi_{i}^{0}(\vec{\rho}_{3})
\phi_{j}^{0}(\vec{\rho}_{4})\right],\label{funt}
\end{eqnarray}
where \textit{N$_{i,j,m}$} is the determinant normalization,
$\chi_{0}(z)$ is the fundamental electron (e) or hole (h) ideal QW
state, $\phi_{i}^{0}(\vec{\rho})$ is a s-like one particle
wavefunction (plane wave for the CM in ideal QWs). The coordinates
$\vec{\rho}_{3}$ and $\vec{\rho}_{4}$ are obtained through the
interchange between electrons 1 and 2 in the relative coordinates
($\vec{\rho}_{1}$, $\vec{\rho}_{2}$).

We limit our basis to the fundamental QW states and s-like
orbitals. Although it is known that they are not sufficient for a
quantitative trion description\cite{Dacal00}, the present choice
retains the main physical results of the defects influence on the
trion states. We chose gaussian one particle wavefunctions to
describe the in-plane motion because of their good trion
description in the absence of defects\cite{Dacal00}. This basis
also helps due to the analytical results for the calculation of
all the contributions on the trion Hamiltonian, including the
defect ones.

The actual form of the interface defects is not known and it
depends on the sample growth conditions. Because of this, we
represent all the defects by a gaussian potential :
\begin{eqnarray}\label{defxm}
V_{def}(e,h)=&V&_{e,h}.Y\left(\frac{L}{2}<z_{e,h}<\frac{L}{2}+\delta\right)
\exp\left[-\left(\frac{\vec{\rho}_{e,h}}{D}\right)^2\right]
\end{eqnarray}
Here, we used electron (e) and hole (h) absolute coordinates,
V$_{e,(h)}$ is the QW confining potentials for electrons (holes),
\textit{Y(z)} is the step function (Y(z)=1 if z$>$0 and Y(z)=0 if
z$<$0) and \textit{L} is the QW width. The defect parameters are
$\delta$, the defect depth in the z direction, and D, the defect
radius in the xy plane.

The charged exciton binding energy is defined as the difference
between the energy of this charged complex and the energy of a
trapped exciton (X$^0$) plus an in-plane free electron (hole), in
the X$^-$ (X$^+$) case.

\section{Results and discussions}

Since we are considering a GaAs/Al$_{0.3}$Ga$_{0.7}$As QW, the
effective parameters used are : m$_e$=0.067m$_0$,
m$_{hz}$=0.377m$_0$, m$_{hxy}$=0.112m$_0$, $\varepsilon$=13.2 for
the well and barrier materials. The conduction band off-set is
0.6.

Figure \ref{enertot} shows the energy gain due to the interface
defect presence as a function of QW width for X$^-$ (squares),
X$^+$ (open circles) and excitons (triangles). The following
defect parameters were considered : D = 250 {\AA} and $\delta$ = 1
GaAs monolayer. Our results show that the X$^-$ has the most
significant energy gain. We present the results for the QW width
range in which the internal degrees of freedom are weakly
affected, i. e., the
\textbf{binding} energy gain is less than 40\verb+%+.
The binding energy is the energy distance between each complex and
the respective continuum. Figure \ref{enerrel} shows the
percentage of binding energy that is gained due to the defect
presence as a function of the QW width. We considered the same
defect parameters as in Fig. \ref{enertot}. As one can see again,
the X$^-$ is the most strongly affected complex by the defect,
mainly in the narrow QW range. This happens because the carrier
wavefunction probability inside the defect is greater for narrow
QWs, particularly for electrons (inset of Fig. \ref{enerrel}). Our
results show that a single monolayer fluctuation is sufficient to
produce a drastic effect on the trion binding energy.

Figure \ref{cm} shows the CM mean radius as a function of the QW
width for the same complexes and defect parameters considered in
Fig. \ref{enertot}. It also shows that the X$^-$ is more strongly
localized by the defect than the exciton. This is in agreement
with the experimental findings of Eytan \textit{et
al}.\cite{Eytan98}. However, they attributed the origin of this
strong localization of charged complexes to fluctuations in the
electrical potential of remote ionized donors. Our results show
that even for strictly structural defects the X$^-$ is more
affected than the exciton.

In conclusion, we presented a simple model to take into account
the interface defects contribution on the trion binding energy.
Our results show that the structural imperfections are more
important in the case of narrow QWs and that the charged exciton
are more strongly localized than the neutral one even in the case
of strictly structural defects. This may explain why the
theoretical results have, in general, better agreement with the
experiments in the wide QW limit\cite{Peeters2000,Dacal00}. Our
results also show that the negative trion is more sensitive to the
structural imperfections than the positive one.

\section{Acknowledgements}

This work is supported by FAPESP (Brazil) and CNPq (Brazil). FAEP
(Brazil), NEDO (France) and SQID (France) are also acknowledged
for financial support.

%

\begin{figure}  
 \caption{X$^-$ (squares), X$^+$ (open circles) and exciton (triangles)
 energy gain due to interface defects presence as a function of QW width.
 The defect parameters are : radius, D = 250 {\AA},and depth, $\delta$ = 1 GaAs monolayer.}
 \label{enertot}
 \end{figure}

 \begin{figure}  
 \caption{X$^-$ (squares), X$^+$ (open circles) and exciton (triangles) relative energy gain
 as a function of QW width. Inset : probability of finding an electron (solid line)
 or a hole (dashed line) inside the region of the defect as a function of
 QW width for an ideal QW. The defect parameters are the same as in Fig. \ref{enertot}}
 \label{enerrel}
 \end{figure}

 \begin{figure}  
 \caption{CM mean radius as a function of QW width for
 X$^-$ (squares), X$^+$ (open circles) and exciton (triangles). The defect
 parameters are the same as in Fig. \ref{enertot}}
 \label{cm}
 \end{figure}

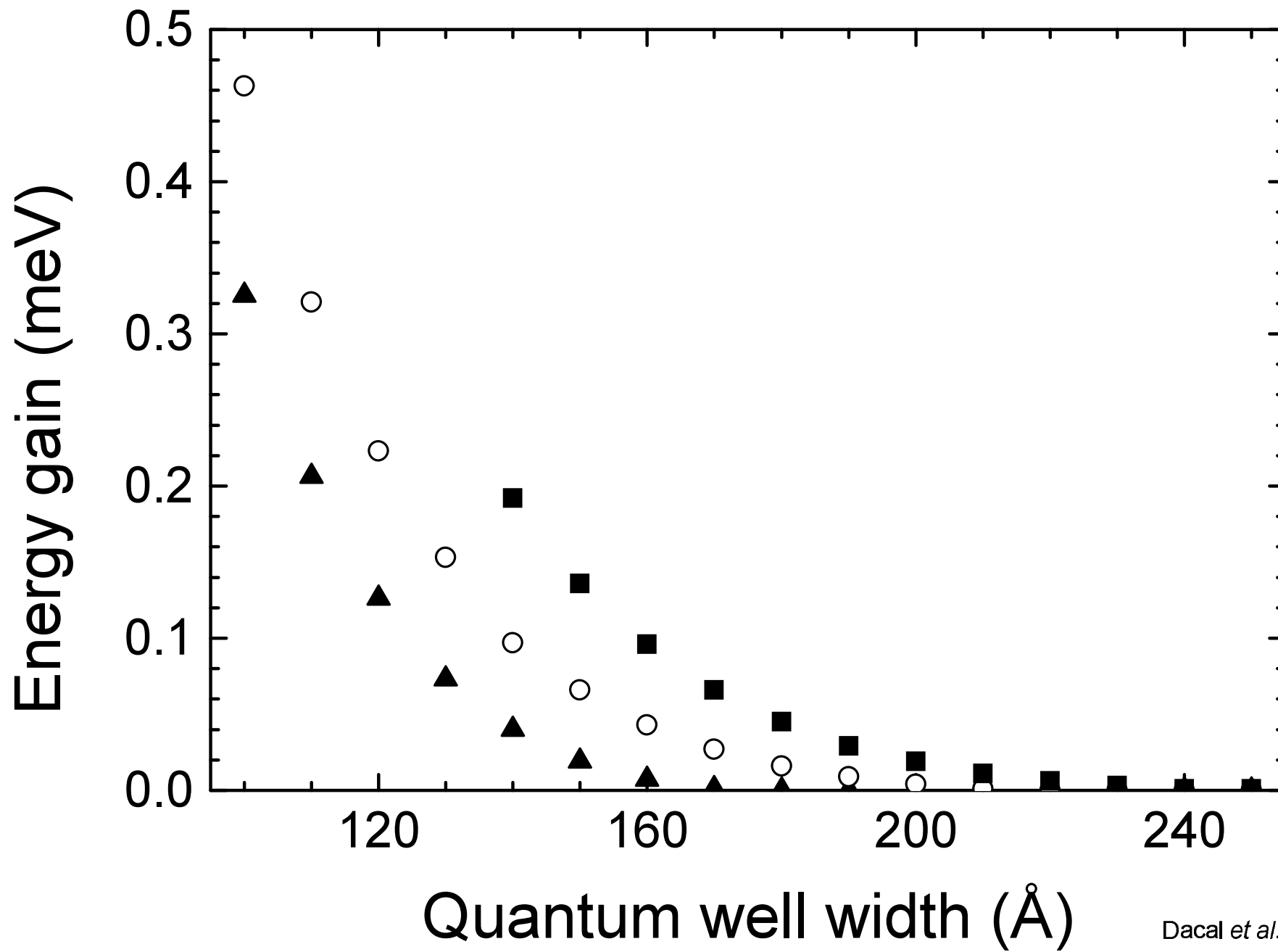
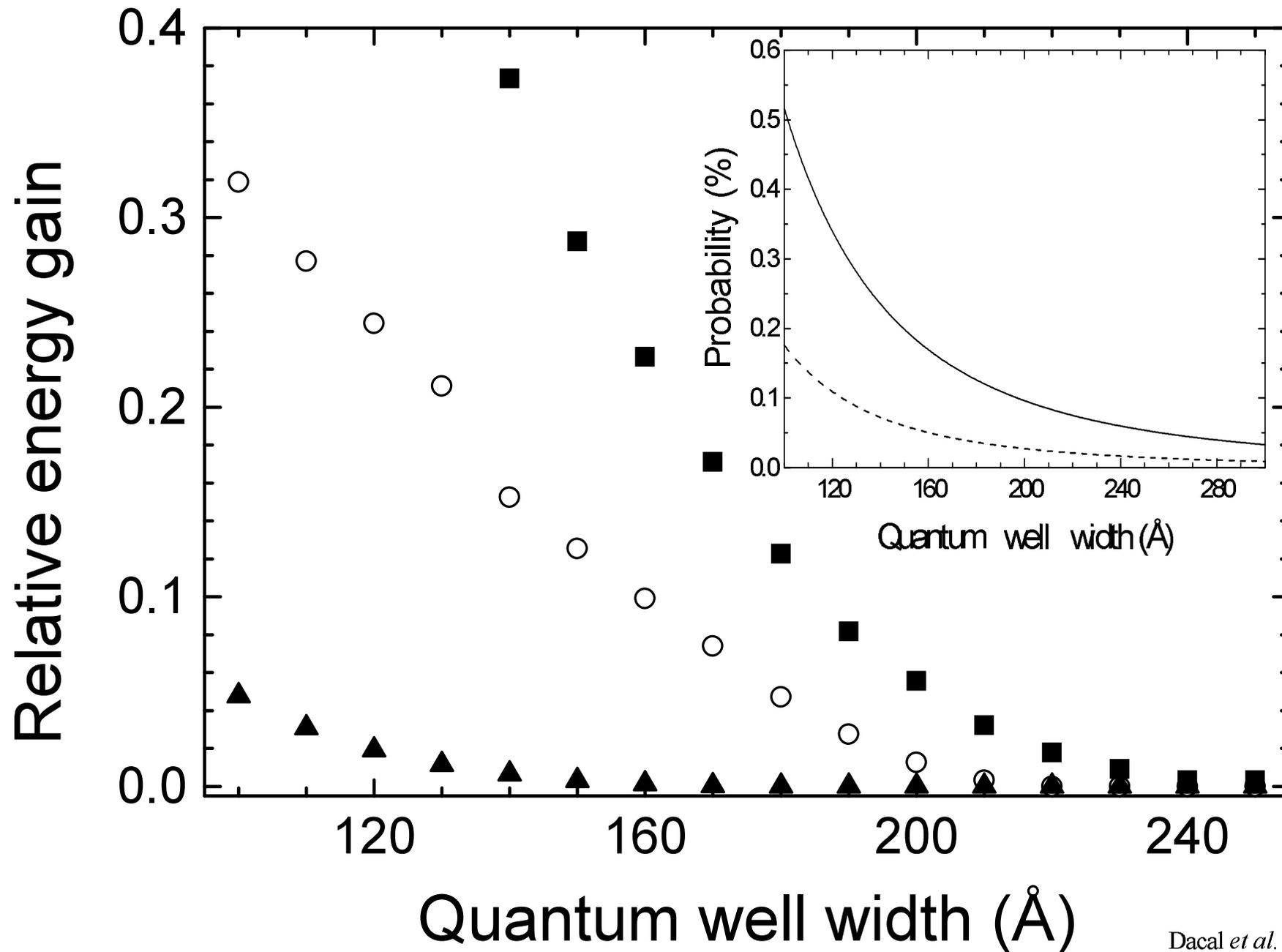
\end{document}